\documentstyle[12pt,fleqn]{article}
\date{}

\title{Finite-Size Effects in Critical Phenomena from Mean Field
Approach with $\phi^4$ Model}
\author{C.B. Yang and X. Cai\\
\footnotesize{Institute of Particle Physics, Hua-Zhong Normal University,
Wuhan 430079, China} }
\begin{document}
\maketitle

\begin{abstract}
The finite-size effects in critical phenomena of a thin film system are
studied from a mean field (MF) approach with $\phi^4$ model for second-order
phase transition. The influence of boundary condition on the critical
properties are emphasized. Scaling functions for relative free-energy
and relative specific heat capacity are given.

\end{abstract}

The critical phenomena in the thermodynamic limit is characterized by
the divergence of the correlation length $\xi$ near critical point. Nowadays,
the experimental techniques have become so advanced that the correlation length
$\xi$ can be pushed up to several thousand \AA\ and the samples under study
become comparable with $\xi$. As a consequence, the effects of finite size of
the samples on the critical phenomena become increasingly important.
Generally such effects depend on the shape of the sample, on the boundary
condition, on the dimension of the system and on the number of components of
the order parameter. Finite-size effects near critical point have been remained
to be a topic of active research over the past two decades. In fully finite or
quasi-one-dimensional systems the phase transition is smeared out, whereas
in thin films of thickness $L$ the critical temperature $T_C(L)$ is shifted
with respect to the bulk $T_C$. Without any doubt,
detailed theoretical studies are necessary in order to understand the most
important finite-size scaling structures predicted by phenomenological [1]
and renormalization-group theories [2].

Since 1985 various perturbative approaches to calculations of finite-size
scaling functions with the $\phi^4$ model have been developed [3-6]. All
those perturbative approaches are based on the division of the free energy
into a Gaussian term and higher order perturbative terms. In fact, the
coupling constant with the $\phi^4$ model of the perturbation is not necessary
small, so that the convergence of the perturbation expansion cannot be
ensured. Thus some more effective approaches to the calculation are needed.
It is clear that the random changes of the order parameter due to thermal
motion should be around certain average and the fluctuations around
the average are generally not too large except in the temperature region
close to the critical one. If the averaged order parameter
can be obtained, the fluctuations can be described as a Gaussian term
and small higher order terms. MF
calculations are important. First of all, they furnish the lowest order
approximation of the real problem, and some characteristics can be obtained
from such calculations. The advantage of the MF approximation is that
it allows a systematic discussion of the corrections to that approximation.
For an infinite size system, such a MF calculation is quite easy
and can show some features of critical phenomena. If the same results are
used for system with finite size, thermodynamic quantities will have the
same jumps as the bulk system, thus are physically unacceptable for finite
size systems. The normal solution to that problem is with the zero-mode
approximation instead of the MF approach. The fundamental problem in the
usual MF approach when used to finite size systems, it seems to us, is that
there exist boundary influences. For system with boundaries, one should
consider the influences of the boundaries on the thermal properties near
the bulk critical point. In addition, the spatial distribution of the order
parameter should be taken into account for finite size system. In the MF
calculations for an infinite system, the gradient term has no contribution,
because the system prefers a uniform distribution of order parameter. For
a finite size system, however, the spatial distribution cannot be considered
as uniform any longer due to the influence of the boundary though the condition
of minimum free energy would prefer a smooth distribution. Thus the gradient
term should not be ignored in the MF calculations of finite size system.
When this term is taken into account for a finite size system, the MF
calculations become much more difficult than those for the
infinite system. On the other hand, MF calculations provide the base for the
consideration of fluctuations. The MF approximation is just the tree level
approximation of the corresponding field theory, and corrections to mean
field can be given by suitable perturbation theories. By the way,
the calculations can be expected to give reasonable results
and are very easy comparing to those in perturbation expansion.
Exact calculations show that the MF approximation becomes exact for an
infinite system
when the dimension $d$ of the system becomes very large.
To the best of our knowledge, no such work for finite size system
has ever been done yet. In this paper, we try to show some
features of finite-size effects in second-order phase transition from a MF
approach suitable for limited systems. For simplicity, we only
consider $\phi^4$ model for a thin film under second-order phase transition
with a one-component order parameter as an example to
show the method. The extension to fully finite system and to the case of
$n$-component order parameter is straightforward.

In Ginzburg-Landau theory of phase transition, the partition function for
second-order phase transition is given as
\begin{equation}
Z=\int {\cal D}\phi\,\exp(-H)
\end{equation}

\noindent with effective Hamiltonian
\begin{equation}
H=\int  d^d\,x\,\left[{\gamma\over 2}\phi^2+{c\over 2}(\bigtriangledown
\phi)^2+{b\over 4!}\phi^4\right]
\end{equation}

\noindent with $\gamma=a(T-T_C),\ \ a>0$.

In the following, we consider a thin film with thickness $L$ and assume the
boundary condition for the system is of Dirichlet type for simplicity. For
non-homogeneous boundary conditions, similar calculations can be done.
The MF is the configuration of $\phi$ which makes the effective Hamiltonian
$H$ minimum, or simply $\delta H=0$. From the point view of partition
function, MF theory is a saddle point approximation. The corrections to the
approximation can be considered by expanding around the saddle point.
From the condition of minimum of $H$, one gets
\begin{equation}
\bigtriangledown^2 \phi=\gamma\phi+{b\over 6}\phi^3
\end{equation}

\noindent with boundary condition $\phi(0)=\phi(L)=0$. Since one intends to
choose the configuration of $\phi$ with $H$ minimum, the averaged order
parameter $\phi$ should vary only in the direction of $L$ which will be chosen
as $x$ direction in later discussions. If $\gamma>0$, last equation has only
trivial solution $\phi=0$, the same as for the bulk system. So one needs only
to discuss the case with $\gamma<0$. Defining $\phi_0=\sqrt{-6\gamma/b}$
and $\xi^2=-1/\gamma$, the last equation can be normalized as
\begin{equation}
\xi^2 {{\rm d}^2\over {\rm d}x^2}\Psi=-\Psi+\Psi^3\ ,
\end{equation}

\noindent with $\Psi=\phi/\phi_0$ and $\Psi(0)=\Psi(L)=0$. Obviously,
$\xi$ is the usual correlation length and $\Psi$ is MF relative
to $\phi_0$, the MF for the corresponding infinite system.
Apparently, Eq. (4) is invariant under $\Psi\leftrightarrow -\Psi$
since $\pm\Psi$ correspond to the same effective Hamiltonian $H$.
Then one can assume $\Psi\geq 0$ in later discussions.
Last equation can be integrated as
\begin{equation}
\xi^2 \left({{\rm d}\Psi\over {\rm d}x}\right)^2=-\Psi^2+{\Psi^4\over 2}+c\ ,
\label{aa}
\end{equation}

\noindent with $c$ an integral constant which, as will be shown afterward,
is in the range [0,1/2]. It should be kept in mind that the solution
of last equation should make the effective Hamiltonian $H$ minimum.
Thus the spatial change
of $\Psi$ should be as slow as possible. Due to the symmetric boundary
condition, $\Psi$ will reach its unique maximum at $x=L/2$, where
${\rm d}\Psi/{\rm d} x=0$. Let $k^2=(1-c-\sqrt{1-2c})/c$, the integral
constant $c$ as a function of $\xi$ and $L$ can be determined through
\begin{equation}
L=2\xi\sqrt{1+k^2}F(k)\ ,
\end{equation}

\noindent where $F(k)$ is the first kind of complete elliptic integrals.
From above expression, one can see that the integral constant $c$ depends
not on $\xi$ and $L$ separately but on the combination of $L/\xi$. Thus it
is convenient to define a reduced length of the system $l=L/\xi=l_0
\sqrt{1-T/T_C}$. The value of $l_0$ depends only on the nature of the
substance under study. In fact both $l$ and $l_0$ are dimension-less, they are
ratios between the real thickness of the film and certain characteristic
lengths. In fact, Eq. [4] can be written in terms of dimensionless variable
$l$. Since $k$ should be in the interval [0, 1] the integral constant $c$
should be in [0, 1/2]. Smaller $c$ corresponds to smaller reduced length $l$.
The behavior of $c$ with $l$ is shown in Fig. 1. A special feature appears
as shown in the figure. If the reduced length $l\leq \pi$, $c=0$ thus the
maximum of the MF or averaged order parameter $\Psi=0$.
In other words, under such condition the system will remain in the disordered
phase despite the bulk critical point has been crossed from above.
One should notice that what plays a role is not the real length $L$ of
the system but the reduced one $l$ which depends also on the deviation from
the bulk critical point. This shows that if the reduced length of the system
is too short, either due the small size of the system or due to its closeness
to the bulk critical point, the influence of the boundary will dominate. This
result is not surprising considering the physical meaning of the correlation
length $\xi$ which can be regarded as an indication of the influence range of
the boundary condition. The minimum values of the reduced length
$l_{min}=\pi$, as a mathematical
result, can be understood physically. If $c$ is very small, the maximum
of $\Psi$ is also very small, so the $\Psi^4$ term in Eq. (\ref{aa}) can be
neglected. Then the solution of Eq. (\ref{aa}) is $\Psi=\sqrt{c}\sin(x/\xi)$.
As discussed above, $\Psi$ should reach its maximum before $x=L/2$, thus
$l=L/\xi\geq \pi$, otherwise $\Psi\equiv 0$. Of course, the existence of
$\Psi^4$ term will modify the behavior of $\Psi$ and make the two parts for
$x>L/2$ and $x<L/2$ join together smoothly for larger $l$.

The existence of a minimum reduced length $l_{\rm min}$ is an
indication of the shift of critical point for a
the finite size system with respect to the bulk one. One
can easily get the critical temperature $T_C(l_0)$ at which the finite system
begins to have non-zero averaged order parameter
\begin{equation}
T_C(l_0)=T_C\cdot\left(1-\left({\pi\over l_0}\right)^2\right)\ .
\end{equation}

\noindent If the characteristic length $l_0$ of the system goes to infinity
(thus the system becomes unlimited), $T_C(l_0)$ in last equation returns to
the bulk value $T_C$. But for small $l_0$ the shift of critical temperature
will be important. Specially, if $l_0\to \pi,\ \ T_C(l_0)\to 0$, so that
the limited system will stay in the disordered phase until the temperature
becomes absolute zero. Thus there exists a lower bound for the size of the
system above which there may exist phase transition. With the increase of
$l$, the constant $c$ increases monotonously and saturates at 1/2 for $l\to
\infty$. When $c\sim 1/2$, the MF $\phi$ almost equals to the value
$\phi_0$ except in the narrow bands near the boundaries. This
shows that the effects of the boundary condition can be ignored for system
with the reduced length $l>>1$. This point can be seen from the behaviors of
other thermodynamic quantities.

The effects of the boundaries on the distribution of MF $\Psi$ can also be
shown from Eq. (\ref{aa}). One can first normalize the film
by variable transformation $x=x^\prime L$. Then
\begin{equation}
x^\prime={1\over l}\sqrt{1+k^2}F(\sin^{-1}{\Psi\sqrt{1+k^2}\over
\sqrt{2}k},k) \ ,
\end{equation}

\noindent where $F(\theta, k)$ is the first kind of incomplete elliptic
integral. From last expression the MF $\Psi$ can be solved as a
function of $x^\prime$ and $l$. Due to the existence of minimum reduced
length $l_{\rm min}=\pi$ one is interested in the varying behavior of the
distribution of $\Psi$ verse $l^\prime\equiv l-\pi$. For $l^\prime$=0.0625,
0.214, 0.328, 0.409, 0.863 and 10.00 the distributions of
MF $\Psi=\Psi(x^\prime, l)$ are shown in Fig. 2. From those figures one
can see that the maximum value of $\Psi$ increases with the increase of $l$
and when $l$ is large enough there is a plateau in the spatial distribution
of $\Psi$. The larger $l$, the larger the
fraction of the plateau, indicating the lesser influence of the boundary
condition. Because of the continuous change of the distribution with the
reduced length, the singularities accompanied with the infinite system are
smeared, as demanded from general physical consideration. This can be shown
from the behaviors of thermodynamical quantities.

To study the finite-size effects on free energy, define $F_\infty=-3
\gamma^2/2b$, which is the density of free-energy for infinite system. Then
the ratio $F_{\rm r}$ between free energy of the thin film per unit thickness
$F/L$ and that of the infinite counterpart per unit volume $F_\infty$
can be shown as
\begin{equation}
F_{\rm r}={F\over L\,F_\infty}={2\over l}\int^l_0 c(t){\rm d}t\ .
\end{equation}

\noindent This ratio depends only on the reduced length of the system,
indicating the mutual influences of the boundaries and intrinsic thermal
motions. From the saturation of $c(l)$, it is obvious that $F_{\rm r}$ also
saturates to 1 for $l\to \infty$. But as shown in Fig. 3, the saturation of
$F_{\rm r}$ to its maximum value 1 is very slow, so to speak, the
influences of the boundary condition is in force even the length scale $L$ of
the system is over ten times larger than the correlation length $\xi$. Thus
the influences of boundaries should be considered even for systems with size
scale much larger than $\xi$.

Once the free-energy of the limited system is known, other thermal quantities
can be obtained. As an example, one can consider the specific heat capacity.
Define $C_{\rm r}$ the ratio between specific heat capacity of the limited
system and that of the infinite counterpart.
After simple algebras, one arrives at
\begin{equation}
C_{\rm r}\equiv {C_V(l)\over C_V(\infty)}
=\frac{5}{4}c+\frac{3}{8}F_{\rm r}+\frac{1}{4}l{{\rm d}\,c\over
{\rm d}\,l} \ .
\end{equation}

\noindent Again this ratio $C_{\rm r}$ only depends on the reduced length of
the system. The vary of $C_{\rm r}$ with the reduced length $l$ are shown in
Fig. 4. In the region of $l$ close to its critical value $\pi$, the relative
specific heat capacity increases quickly with $l$, but the rate of increase
slows down and finally saturates to 1 as expected.

As a summary, we have investigated the finite-size effects on critical
behaviors based on a MF approach adapted for the limited system with a
one-component order parameter. Though the fluctuations have not
been considered, interesting critical phenomena due to finite size of
the system have been shown. The relative thermodynamical quantities
such as the free energy and specific heat capacity with respect to those
for infinite counterpart are expressed as scaling functions
of the reduced length $l$. It will be fruitful to study the modification to
the MF results from fluctuations, which will be discussed elsewhere.

\vskip 1cm
\section*{References}

\begin{description}
\item{[1]} M.E. Fisher, in: {\it Critical Phenomena, International School of
Physics ``Enrico Fermi''}, Course 51, ed. M.S. Green
(Academic, New York, 1971).

\item{[2]} M. Suzuki, Prog. Theor. Phys. {\bf 58}(1977)1142.

\item{[3]} E. Br\'{e}zin and J. Zinn-Justin, Nucl. Phys. {\bf B257}(1985)867.

\item{[4]} J. Rudnick, H. Guo and D. Jasnow, J. Stat. Phys. {\bf 41}(1985)353.

\item{[5]} A. Esser, V. Dohm and X.S. Chen, Physica {\bf A222}(1995)355;
X.S. Chen, V. Dohm and N. Schultka, Phys. Rev. Lett. {\bf 77}(1996)3641.

\item{[6]} C.B. Yang and X. Cai, cond-mat/9612074.

\end{description}

\vskip 2cm
\begin{center}
\section*{Figure Captions}
\end{center}
\begin{description}
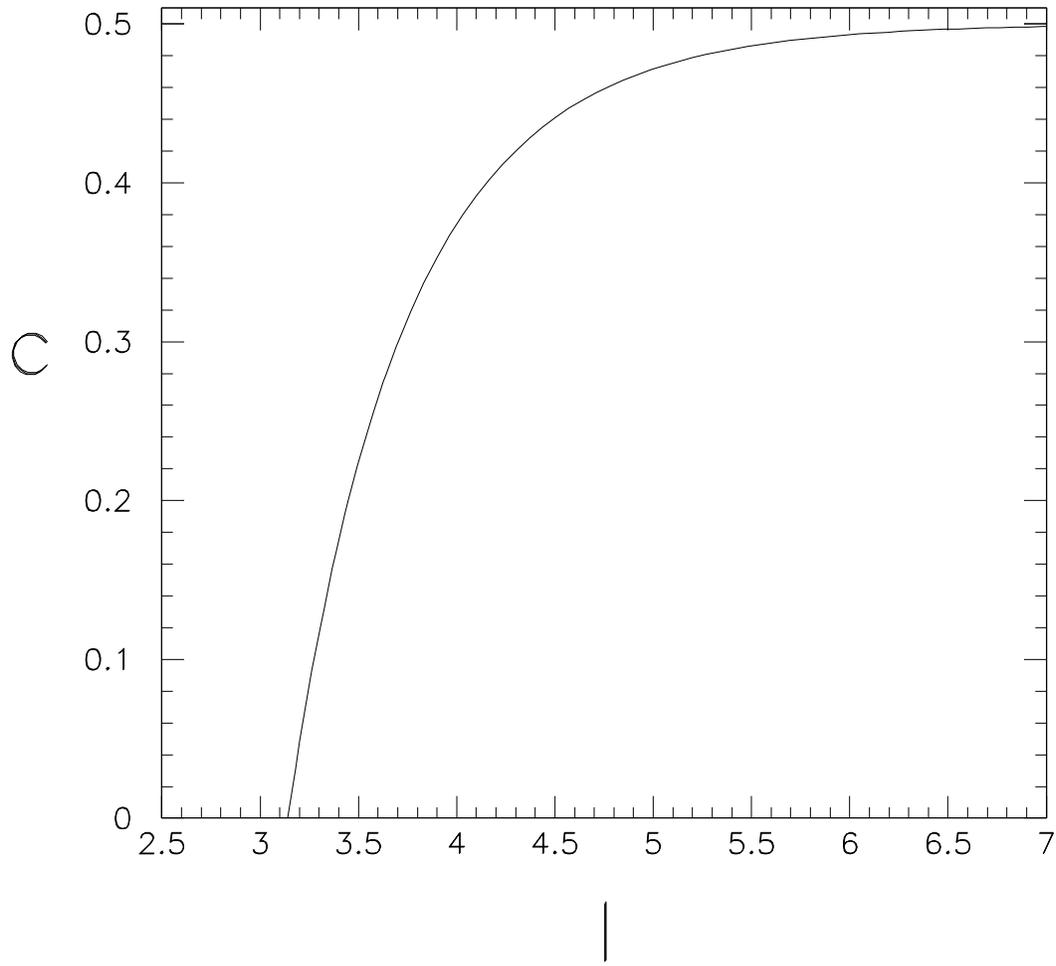
\item {\bf Fig. 1} \ \ The behavior of the integral constant $c$ as a function
of the reduced length $l$ of the system.

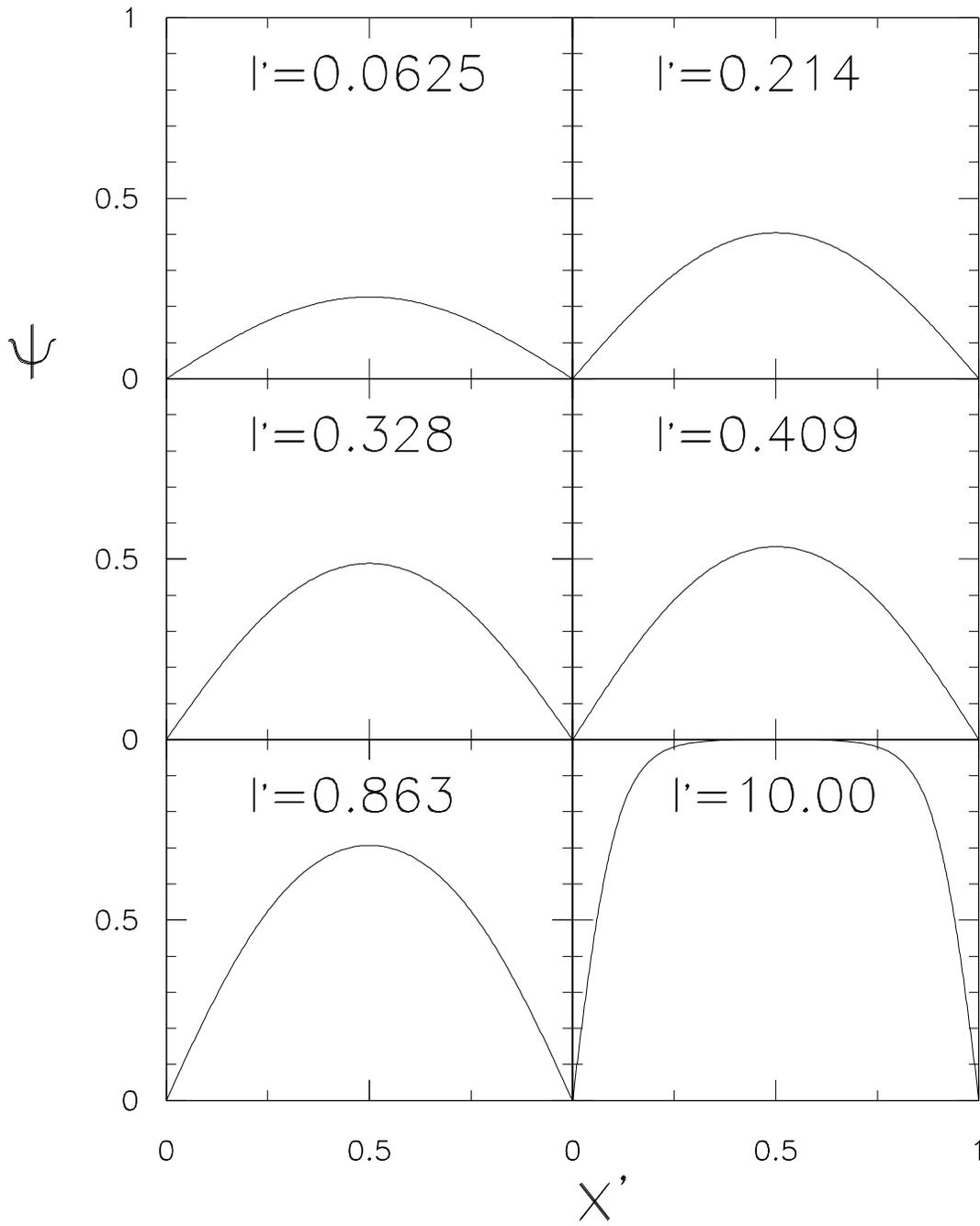
\item {\bf Fig. 2} \ \ Distributions of mean field $\Psi$ as defined in text
over the normalized space $x^\prime$ for various $l^\prime=l-\pi$.

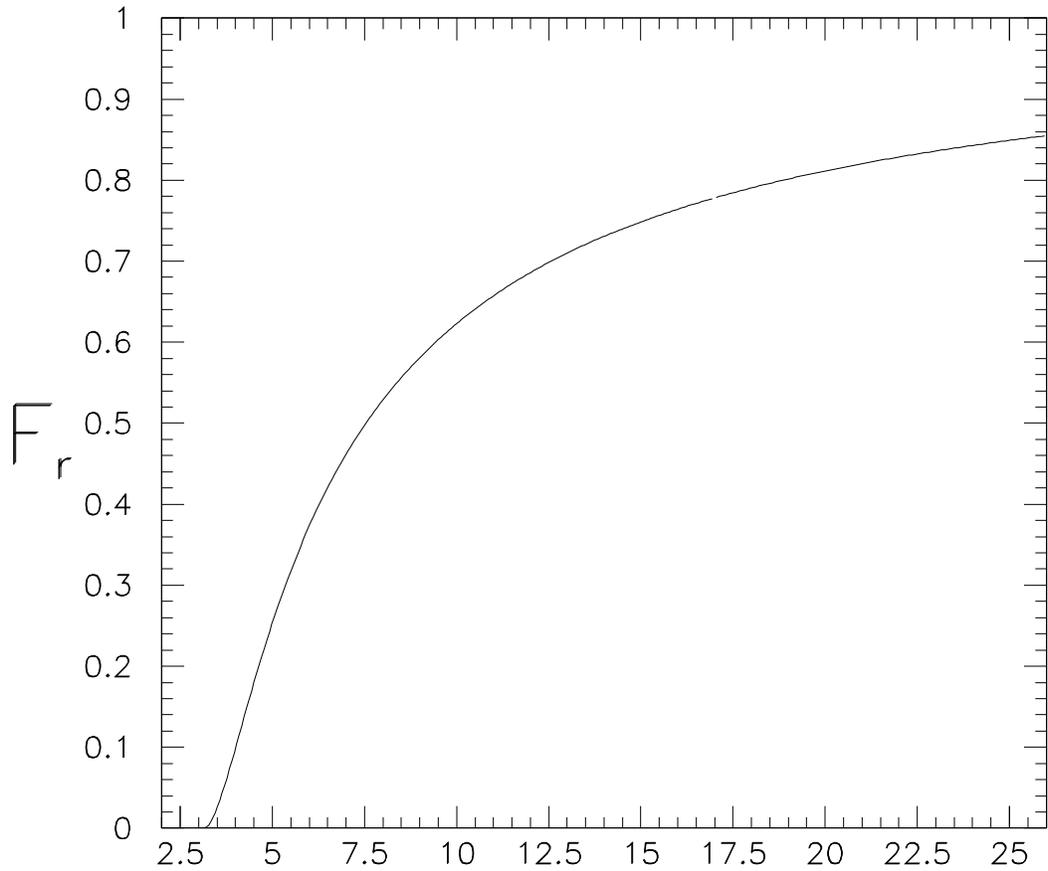
\item {\bf Fig. 3}\ \
The ratio between the free energy density of a limited
system and that of the infinite counterpart as a function of the reduced
length of the system.

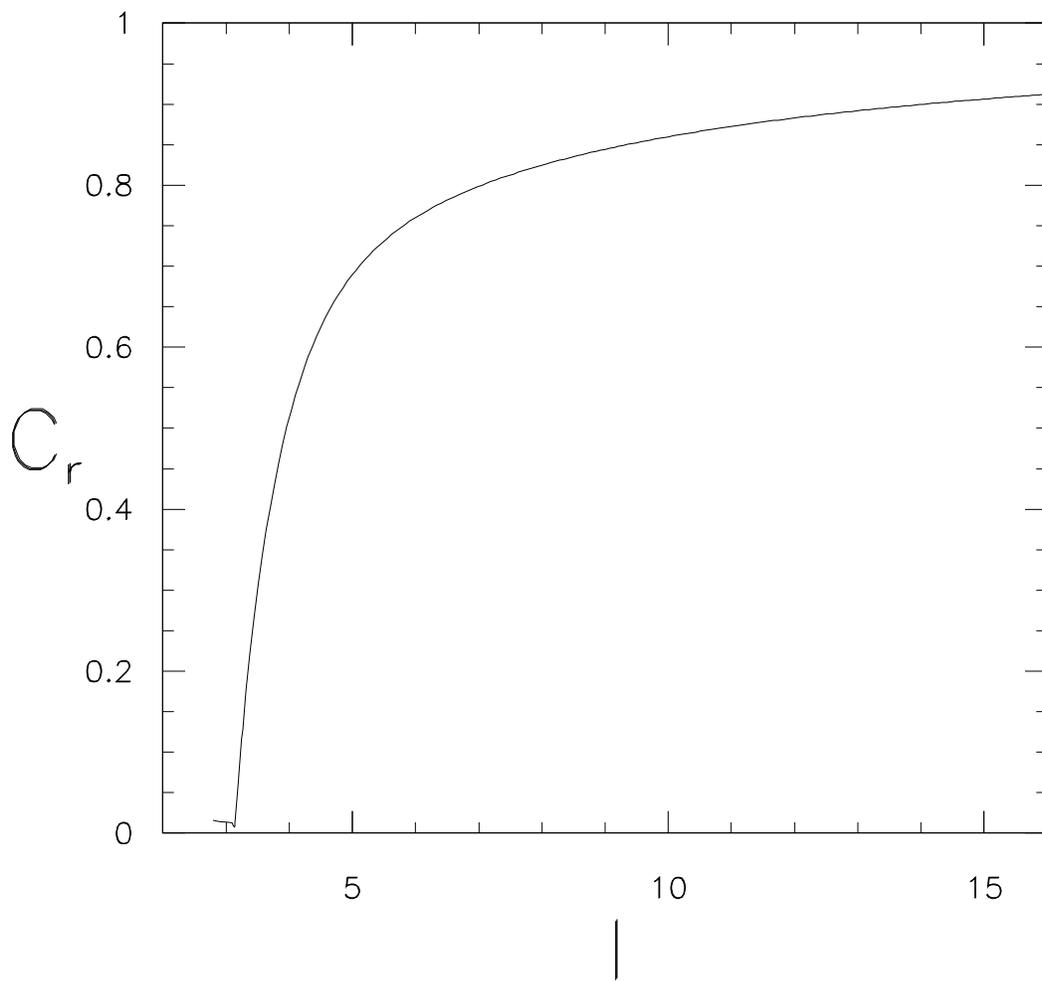
\item {\bf Fig. 4}    \ \
The ratio between the specific heat capacity of a limited
system and that of the infinite counterpart as a function of the reduced
length of the system.
\end{description}

\end{document}